\documentclass[prb,showpacs,preprint]{revtex4}
\usepackage{amsmath,graphicx}

\def\prl{Phys. Rev. Lett.}
\def\prb{Phys. Rev. B}

\begin{document}

\title{Photonic band mixing in linear chains of optically coupled micro-spheres}

\author{L. I. Deych}
\author{A. Roslyak}

\affiliation{Physics Department, Queens College, City University of
New York, Flushing, New York 11367, USA}

\begin{abstract}
The paper deals with optical excitations arising in a
one-dimensional chain of identical spheres due optical coupling of
whispering gallery modes (WGM). The band structure of these
excitations depends significantly on the inter-mixing between WGMs
characterized by different values of angular quantum number, $l$. We
develop a general theory of the photonic band structure of these
excitations taking these effects into account and applied it to
several cases of recent experimental interest. In the case of bands
originating from WQMs with the angular quantum number of the same
parity, the calculated dispersion laws are in good qualitative
agreement with recent experiment results. Bands resulting from
hybridization of excitations resulting from whispering gallery modes
with different parity of $l$ exhibits anomalous dispersion
properties characterized by a gap in the allowed values of
\emph{wave numbers} and divergence of group velocity.
\end{abstract}


\maketitle
\section{Introduction}
\label{intro} Recent proposal of coupled resonator optical
waveguides  and optical filters\cite{YarivOptLett1999,YarivQE2002}
stimulated interest in systems of optically coupled micro-spheres.
It has been known for a long time that electromagnetic modes of an
individual sphere (Mie resonances) with large enough values of their
angular momentum can have very long radiative
life-times\cite{book:Opt_Proc_MC}. These are so called whispering
gallery modes (WHM) characterized by the concentration of the
electromagnetic field along the surface of a sphere with an
evanescent tail escaping outside. Such field configuration makes it
possible to optically couple two spheres positioned in the proximity
of one another. Initial work on the optical coupling of the
micro-spheres was concentrated on the case of just two spheres (the
arrangement is known as photonic atoms or molecules), where the
splitting of the modes and the formation of the coupled states was
observed\cite{ArnoldJOSA1992,MukaiyamaPRL1999,RakovichPRA2004}.
Recently, however, an interest has shifted toward linear chain of
many spheres, which  are envisioned as building blocks of various
photonic devices, such as waveguides\cite{MoellerOL2005} or delay
lines\cite{MukaiyamaPRL2005}. The obvious result of optical coupling
in this system is formation of collective optical excitations
propagating along the chain, which we will call \emph{supermodes} in
order to distinguish them from the modes of individual spheres. The
supermodes form photonic bands, which were observed by several
research groups almost at the same
time\cite{AstratovAPL2004,MukaiyamaPRL2005,MoellerOL2005}.

One of the popular tools used to analyze experimental
results\cite{MukaiyamaPRL2005} is a simple phenomenological
dispersion law of a tight-binding type
\begin{equation}\label{eq:SMTBMdisp_eq}
\omega=\omega_0+\kappa\cos(qd)
\end{equation}
where $\omega$ and $\omega_0$ are respectively the frequency of the
supermode, characterized by a wave number $q$, and that of a single
mode WGM resonance; phenomenological coefficient $\kappa$
characterizes the strength of the optical coupling, and $d$ is the
period of the structure. This approach, which was originally
suggested in Ref.\onlinecite{YarivOptLett1999} to describe modes of
coupled cavities, derive every photonic band from a respective
single sphere WGM resonance, characterized by its angular momentum,
$l$. Its applicability is based on the assumption that the modes of
coupled spheres with different $l$ do not mix, which is certainly
not true in the exact sense, but can be approximately valid or not
valid at all depending upon the type of modes under consideration.

The admixture of WGM with different angular momentums arises already
for two interacting spheres, and results from the violation of
spherical symmetry in such systems. It has been realized early on
that the inter-mode coupling is primary responsible for radiative
decay of the coupled modes\cite{FullerApplOpt1991}, and may also
affect positions of the resonances in a bi-spherical structure. We
will show in this paper that these effects are even more important,
and in certain situations, crucial, for the supermodes of the
multi-spherical chains. In these systems, instead of individual
resonances, one has to consider mixing between bands of collective
propagating excitations. Using analogy with solid state physics one
can call this phenomenon band-mixing or band hybridization. In this
paper, we present a theory of the inter-band coupling in linear
chain of spheres based on the tight-binding approach to the optical
coupling. This approach was originally formulated in
Ref.\onlinecite{SoukoulisPRL1998}, and carefully analyzed for the
case of bi-spheres in Ref.\onlinecite{MiyazakiPRB2000}.

Our main goal in this paper is to extend the analysis of
Ref.\onlinecite{MiyazakiPRB2000} to a linear chain of spheres, and
apply it to the problem of photonic band structure and dispersion
laws of the quasi-stationary (long-living) collective optical
excitations of the chain. Using a perturbative approach, we derive
analytically dispersion equations characterizing these excitations
and analyze them in several particular cases of recent experimental
interest. We show, in particular, that in spite of a formal analogy
between optical and electronic tight-binding models, the photonic
band mixing may lead to very unusual effects, which do not have
counterparts in solid state systems.

\section{Tight-binding model for a linear chain of dielectric
spheres}

\begin{figure}
  \includegraphics[width=3in]{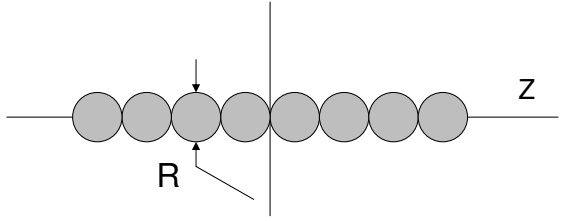}\\
  \caption{The chain of spheres of radius $R$ with $Z$-axis along the axis of the chain}\label{fig:spheres}
\end{figure}

The system, which will be considered in the paper consists of a
number of identical spheres with radius $R$ and refractive index
$n$, whose centers are all aligned along the same line (see
Fig.\ref{fig:spheres}) at a distance $d\geq 2R$ from each other. We
will be interested in collective excitations of this chain, which in
the spirit of tight-binding approach is described as a combination
of modes of individual spheres. It is pertinent, therefore, to start
with a brief review of the properties of a single sphere.
Electromagnetic field in this case can be described by vector
spherical harmonics (VSH), whose angular dependence is specified by
angular number $l$ (angular momentum) and azimuthal number $m$
(z-component of the angular momentum). The radial part of the VSH,
$f_l(kr)$ is given either by spherical Hankel function of the first
kind, $h_l^{(1)}(kr)$, for points outside of the sphere or by
spherical Bessel function, $j_l(nkr)$,  for points inside the
sphere. Parameter $k$ here represents vacuum wave number,
$k=\omega/c$, where $\omega$ and $c$ are frequency and the speed of
light in vacuum. We will use notation $\mathbf{N}_{m,l}(\mathbf{r})$
and $\mathbf{M}_{m,l}(\mathbf{r})$ for VSH describing
electromagnetic waves of TE and TM polarizations respectively.
Explicit expressions for functions $\mathbf{N}_{m,l}(\mathbf{r})$
and $\mathbf{M}_{m,l}(\mathbf{r})$ can be found in many sources
(see, for instance, Ref.\onlinecite{stratton_book1941}) thus we do
not present it here. An isolated dielectric sphere is an open
system, therefore, it is more prudent to describe its properties in
terms of resonances rather than normal modes. To this end one
presents an incident, $\mathbf{E_{inc}}$, and scattered by the
sphere, $\mathbf{E_{s}}$, waves in terms of VSH:
\begin{eqnarray}
\mathbf{E_{inc}}&=&\sum_{l,m}\left[\zeta_{l,m}\mathbf{N}_{m,l}(\mathbf{r})+\eta_{l,m}\mathbf{M}_{m,l}(\mathbf{r})\right]\label{eq:inc_wave_expansion}\\
\mathbf{E_{s}}&=&\sum_{l,m}\left[a_{l,m}\mathbf{N}_{m,l}(\mathbf{r})+b_{l,m}\mathbf{M}_{m,l}(\mathbf{r})\right]\label{eq:sc_wave_expansion}.
\end{eqnarray}
Using Maxwell boundary conditions one establishes relations between
coefficients $a_{l,m}$ and $b_{l,m}$ of
Eq.(\ref{eq:sc_wave_expansion}) and coefficients $\zeta_{l,m}$ and
$\eta_{l,m}$ of Eq.(\ref{eq:inc_wave_expansion}) in the following
form:
\begin{eqnarray}
a_{l,m}&=&\alpha_{l,m}\zeta_{l,m}\label{eq:single_sphereTE}\\
b_{l,m}&=&\beta_{l,m}\eta_{l,m}\label{eq:single_sphereTM},
\end{eqnarray}
where scattering coefficients $\alpha$ and $\beta$ are given by
\begin{eqnarray}
\alpha_{l,m} &=&\dfrac{n^2 j_{l,m} \left(  nx\right)  \left[
xj_{l,m} \left( x\right)  \right]  ^{\prime} -j_{l,m} \left(
x\right) \left[ nxj_{l,m} \left( nx\right)  \right]  ^{\prime}
}{n^{2} j_{l,m} \left( nx\right)  \left[ xh_{l,m}^{\left(  1\right)
} \left( x\right) \right]  ^{\prime} -h_{l,m}^{\left(  1\right)  }
\left( x\right) \left[  nxj_{l,m} \left( nx\right)  \right]
^{\prime} }\label{eq:alpha}\\
\beta_{l,m}&=&\dfrac{j_{l,m}\left(  nx\right)  \left[ xj_{l,m}\left(
x\right)  \right]  ^{\prime}-n^{2}j_{l,m}\left( x\right)  \left[
nxj_{l,m}\left(  nx\right)  \right] ^{\prime}}{j_{l,m}\left(
nx\right) \left[
xh_{l,m}^{\left(  1\right)  }\left(  x\right)  \right]  ^{\prime}-n^{2}%
h_{l,m}^{\left(  1\right)  }\left(  x\right)  \left[ nxj_{l,m}\left(
nx\right) \right]  ^{\prime}}\label{eq:beta}
\end{eqnarray}
Here we introduced the dimensionless parameter $x$ defined as
$x=kR$; $f^\prime (x)$ means a derivative of $f$ with respect to
$x$. In what follows we will use $x$ as a dimensionless  frequency.
The coefficients $\alpha$ and $\beta$ as functions of $x$ possess
poles on a complex plane. The real parts of these poles  give
frequencies of the respective scattering resonances, while the
imaginary parts determine the width of the resonances. It is
customary to characterize the latter using a quality factor $Q$
defined as $Q=\operatorname*{Re} x/\operatorname*{Im} x$. The
resonances are characterized by polarization (TE or TM, depending on
their origin from $\alpha$ or $\beta$ coefficients, respectively),
and two indices, $l$, and $s$, where the former is the respective
angular momentum, while the latter is a radial quantum number
enumerating poles of $\alpha$ and $\beta$ with the same values of
$l$, but different radial distribution of the field (the resonances
are degenerate with respect to index $m$). Following
Ref.\onlinecite{MiyazakiPRB2000} we will label resonances as $lTEs$
or $lTMs$. $Q$-factor of these resonances strongly depends on $l$
reaching very high (up to $10^9$) values for large enough
$l$\cite{book:Opt_Proc_MC}. At the same time, for modes with larger
$s$ $Q$ decreases, sometimes by an order of magnitude.

The resonances with high $Q$ values correspond to so called
whispering gallery modes (WGM), and can be sensibly described as
almost stationary (long-living) normal modes of a sphere. The
collectivization of these modes in a chain of spheres is the main
focus of this paper.  The electric field of these modes has an
evanescent character in the vicinity of the surface of their
respective spheres making the tight-binding description of the
optical coupling between adjacent spheres rather
accurate\cite{MiyazakiPRB2000}. This description is based on
generalizing  the expansions for scattered wave given by
Eq.~(\ref{eq:sc_wave_expansion}) to the case of multiple spheres:
\begin{equation}\label{eq:mult_sc_wave_expansion}
    \mathbf{E_{s}}=\sum_{i,l,m}\left[a_{l,m}^{i}\mathbf{N}_{m,l}(\mathbf{r}-\mathbf{r}_i)+b_{l,m}^{i}\mathbf{M}_{m,l}(\mathbf{r}-\mathbf{r}_i)\right].
\end{equation}
where $\mathbf{r}_i$ is the position vector of the center of $i$-th
sphere. The problem is, however, that VSH in this expansion are
defined in different coordinate systems associated with the center
of each sphere. In order to apply boundary conditions one need to
rewrite them in a common coordinate system. This is achieved with
the help of addition
theorem\cite{SteinApplMath1961,CruzanApplMath1962}, which presents
VSH centered at point $\mathbf{r}_j$ in terms of VSH centered at a
point $\mathbf{r}_i$:
\begin{equation}
\begin{split}
\mathbf{N}_{l,m}(\mathbf{r}-\mathbf{r}_j)=\sum_{l^\prime=1}^\infty\sum_{m^\prime=-l^\prime}^{l^\prime}\left[A^{l^\prime,m^\prime}_{l,m}(\mathbf{r}_j-\mathbf{r}_i)\mathbf{N}_{l^\prime,m^\prime}(\mathbf{r}-\mathbf{r}_i)+B^{l^\prime,m^\prime}_{l,m}(\mathbf{r}_j-\mathbf{r}_i)\mathbf{M}_{l^\prime,m^\prime}(\mathbf{r}-\mathbf{r}_i)\right]\\
\mathbf{M}_{l,m}(\mathbf{r}-\mathbf{r}_j)=\sum_{l^\prime=1}^\infty\sum_{m^\prime=-l^\prime}^{l^\prime}\left[A^{l^\prime,m^\prime}_{l,m}(\mathbf{r}_j-\mathbf{r}_i)\mathbf{M}_{l^\prime,m^\prime}(\mathbf{r}-\mathbf{r}_i)+B^{l^\prime,m^\prime}_{l,m}(\mathbf{r}_j-\mathbf{r}_i)\mathbf{N}_{l^\prime,m^\prime}(\mathbf{r}-\mathbf{r}_i)\right],
\end{split}
\end{equation}
where $A^{l,m}_{l^\prime,m^\prime}(\mathbf{r}_j-\mathbf{r}_i)$ and
$B^{l,m}_{l^\prime,m^\prime}(\mathbf{r}_j-\mathbf{r}_i)$ are the so-
called translation coefficients, which depends on the mutual
position of the spheres. Expressions for the translation
coefficients can be found, for instance, in
Ref.\onlinecite{FullerApplOpt1991} or
Ref.\onlinecite{MiyazakiPRB2000}. The latter paper gives a rather
detailed description of methods for computing these coefficients,
which is not an easy computational task, since for modes with large
$l$ it involves factorials of very large numbers. Using the addition
theorem one can derive a system of equations for the expansion
coefficients $a_{l,m}^{i}$ and $b_{l,m}^{i}$\cite{FullerApplOpt1991}
\begin{eqnarray}
a_{l,m}^{i}&=&\alpha_{l,m}\left\{\zeta_{l,m}+\sum\limits_{j\neq i}\sum\limits_{l^\prime,m^\prime}\left[a_{l^\prime,m^\prime}^{j}A_{l,m}%
^{l^\prime,m^\prime}\left(  \mathbf{r}_j-\mathbf{r}_i\right)  +b_{l^\prime,m^\prime}^{j}B_{l,m}^{l^\prime,m^\prime}\left(  \mathbf{r}_j-\mathbf{r}_i\right)\right]\right\}\label{eq:a_coeff_expan} \\
b_{l,m}^{i}&=&\beta_{l,m}\left\{\eta_{l,m}+\sum\limits_{j\neq i}\sum\limits_{l^\prime,m^\prime}\left[b_{l^\prime,m^\prime}^{j}A_{l,m}%
^{l^\prime,m^\prime}\left( \mathbf{r}_j-\mathbf{r}_i\right)
+a_{l^\prime,m^\prime}^{l}B_{l,m}^{l^\prime,m^\prime}\left(\mathbf{r}_j-\mathbf{r}_i\right)\right]\right\}\label{eq:b_coeff_expan}
\end{eqnarray}
Indexes $i$ and $j$ in these equations enumerate spheres, while $l$,
$l^\prime$, $m$, $m^\prime$ correspond to different resonances of
the individual spheres. The structure of these equations can be
understood by noting that it is equivalent to
Eq.~(\ref{eq:single_sphereTE}) and (\ref{eq:single_sphereTM}), in
which  the incident field, characterized by coefficients
$\zeta_{l,m}$ and $\eta_{l,m}$ is replaced by its sum with the field
scattered by all other spheres. This system of equations is valid
for an arbitrary collection of spheres, as indicated in
Ref.\onlinecite{FullerApplOpt1991}. For a linear chain, in which all
position vectors $\mathbf{r}_i$ can be chosen parallel to each other
and to the axis of the chain, these equations can be simplified. If
one chooses the $Z$-axis of the coordinate system along the chain,
the translation coefficients can be shown to become $A_{l,m}%
^{l^\prime,m^\prime}=A_l^{l^\prime}\delta_{m,m^\prime}$ and $B_{l,m}%
^{l^\prime,m^\prime}=B_l^{l^\prime}\delta_{m,m^\prime}$. Thus, the
sum over $m^\prime$ in Eq.~(\ref{eq:a_coeff_expan}) and
(\ref{eq:b_coeff_expan}) disappears, making the component of the
angular momentum along the axis of the chain a conserving quantity.
This fact obviously reflects the axial symmetry of this system. In
what follows, we, for concreteness, shall assume that $m=1$; results
for other values of $m$ can be obtained from our general formulas by
recalculating parameters $\alpha$ and $\beta$ as well as the
translational coefficients. Thus, we can abridge our notations by
dropping index $m$ all together. At this point we also omit the
terms $\eta_{l,m}$ and $\zeta_{l,m}$ describing the external
incident wave, which leaves us with a system of homogeneous linear
equations and the problem of finding their normal modes instead of
the problem of scattering of an external wave.

Thus, we present the equations for the scattering coefficients in
the following form:
\begin{eqnarray}
a_{l}^{i}&=&\alpha_{l}\sum\limits_{j\neq i}\sum\limits_{l^\prime}\left[a_{l^\prime}^{j}A_{l}%
^{l^\prime}\left(  i,j\right)  +b_{l^\prime}^{l}B_{l}^{l^\prime}\left(  i,j\right)\right] \label{eq:linear_chain_a_coeff}\\
b_{l}^{i}&=&\beta_{l}\sum\limits_{j\neq i}\sum\limits_{l^\prime}\left[b_{l^\prime}^{j}A_{l}%
^{l^\prime}\left( i,j\right) +a_{l^\prime}^{j}B_{l}^{l^\prime}\left(
i,j\right)\right],\label{eq:linear_chain_b_coeff}
\end{eqnarray}
where the position vectors, $\mathbf{r}_i$, of the spheres in the
arguments of the translation coefficients are for shortness replaced
with sphere's numbers, $i$. Eq.(\ref{eq:linear_chain_a_coeff}) and
(\ref{eq:linear_chain_b_coeff}) have been used by Miyazaki and Jimba
in Ref.\onlinecite{MiyazakiPRB2000} for exact numerical analysis of
a bi-sphere. They also developed for this system  an approximate
method of solution of these equations that reproduced results of the
exact calculations with a good accuracy. Here we generalize this
method for a system with an arbitrary, including infinite, number of
spheres, where its use  becomes crucial since exact numerical
calculations grow increasingly more involved and expensive with the
increase in the number of spheres in the chain. We will show here
that using the expanded version of the tight-binding approximation
we are able to derive analytically dispersion equations describing
collective excitations of the chain, which can be solved numerically
with minimal computational efforts.

The reduction of Eq.(\ref{eq:linear_chain_a_coeff}) and
Eq.(\ref{eq:linear_chain_b_coeff}) to an analytically tractable form
is based on several important properties of the translation
coefficients, $A_{l}^{l^\prime}$, and $B_{l} ^{l^\prime}$. First of
all, in the case of $l\gg 1$ the calculation of these coefficients
can be significantly simplified with the help of the so-called
maximum term approximation\cite{MiyazakiPRB2000} that allows
presenting these coefficients in the following form
\begin{eqnarray}
A_{l}^{l^\prime} \left(  i,j\right) & \cong&-2l\left(  -1\right)
^{l+1} \sqrt{\dfrac{l+l^\prime}{\pi\left(  l^\prime+1\right)  \left(
l-1\right) } } \times \dfrac{l^{l} {(l^{\prime})}^{l^\prime}
}{\left(l^\prime+1\right)^{l^\prime+1} \left(
l-1\right)^{l-1} }h_{l+l^\prime}^{(1)} \left(  \eta x\left|  i-j\right|  \right) \label{eq:A_max_term}\\
B_{l}^{l{^\prime}} \left(  i,j\right) & \cong &i\dfrac{x\left|
i-j\right|}{ll^\prime}A_{l}^{l{^\prime}}\left(i,j\right)
\label{eq:B_max_term},
\end{eqnarray}
where $\eta$ defined as $\eta=|\mathbf{r}_i-\mathbf{r}_j|/R\ge 2$ is
a dimensionless distance between the centers of the spheres. In the
case of spheres touching each other, which we shall assume in our
numerical calculations, $\eta =2$.  Taking into account properties
of the Hankel function $h_l^{(1)}(x)$ in Eq.(\ref{eq:A_max_term}),
one can see that the translation coefficients quickly decrease with
the distance between the spheres. This property, which is a
manifestation of the evanescent nature of the optical coupling
between the spheres, allows one to keep in the sum over the spheres
in Eq.(\ref{eq:linear_chain_a_coeff}) and
(\ref{eq:linear_chain_b_coeff}) only terms with $j=i\pm 1$. The
resulting equations constitute the nearest neighbors approximation
for the chain of the spheres. The sum over $l^\prime$ describes
coupling between supermodes with different angular modes, which is
the main subject of the present paper. These equations also contain
terms proportional to translation coefficients $B_l^{l^\prime}$,
which are responsible for coupling between supermodes with different
polarizations. It follows, however, from Eq.(\ref{eq:B_max_term})
that for the nearest neighbors $B_{l}^{l{^\prime}}\ll
A_{l}^{l{^\prime}}$ for $l,l^\prime\gg 1$. We will see later that
the main contribution to the sum over $l^\prime$ comes mostly from
$l^\prime\lesssim l$, and the cross-polarization coupling can,
therefore, be neglected in most cases. In what follows we will also
discard terms $\zeta_l$ and $\eta_l$ responsible for external
excitation. This leaves us with a system of homogeneous linear
equations and the problem of finding their normal modes instead of
the problem of scattering of an external wave. Thus, we present the
final tight-binding equations describing $TE$ and $TM$ supermodes as
\begin{eqnarray}
\frac{1}{\alpha_{l}}a_{l}^{i}&=&\sum\limits_{l^\prime}A_{l}^{l^\prime}\left(a_{l^\prime}^{i-1}  +a_{l^\prime}^{i+1}\right) \label{eq:TE_NNA}\\
\frac{1}{\beta_{l}}b_{l}^{i}&=&\sum\limits_{l^\prime}A_{l}^{l^\prime}\left(b_{l^\prime}^{i-1}+b_{l^\prime}^{i+1}\right),\label{eq:TM_NNA}
\end{eqnarray}
where the translation coefficients
$A_{l}^{l^\prime}=A_{l}^{l^\prime}(i,i+1)$ play the role of the
inter-site coupling parameters of the tight-binding approximation.
They also play an important additional role of being coupling
coefficients for supermodes with different $l$. These equations are
very similar to electronic tight binding equations describing, for
instance, s-p
hybridization\cite{book:Harrison,VOGLTightBinding1983}. There is,
however, a significant difference between optical and electronic
problems caused by the non-Hermitian nature of matrix
$A_{l}^{l^\prime}$ describing the coupling in the optical case. We
will see below that this peculiarity of optical tight-binding
equations is responsible for the anomalous dispersion properties of
the system under consideration. The non-Hermitian nature of the
optical band coupling appears also in a multiple scattering (optical
Korringa-Kohn-Rostocker) approach to this problem developed in
Ref.\onlinecite{MorozMSTPRB1995}.
\section{Collective excitations in the linear chain of micro-spheres}
\subsection{Single band approximation}\label{sec:single_band}
Neglecting terms with $l^\prime\ne l$ in equations (\ref{eq:TE_NNA})
and (\ref{eq:TM_NNA}) one obtains a simple single-band
nearest-neighbor interaction model describing bands of the
collective supermodes originating from $lTEs$ or $lTMs$ WGM's of an
individual sphere with frequencies $x_{ls}$. For a TE band, for
instance, one has
\begin{equation}
\frac{1}{\alpha_{l}}a_{l}^{i}=A_{l}^{l}\left(a_{l}^{i-1}
+a_{l}^{i+1}\right) \label{eq:TE_SM}
\end{equation}
\begin{figure}
  \includegraphics[width=5in]{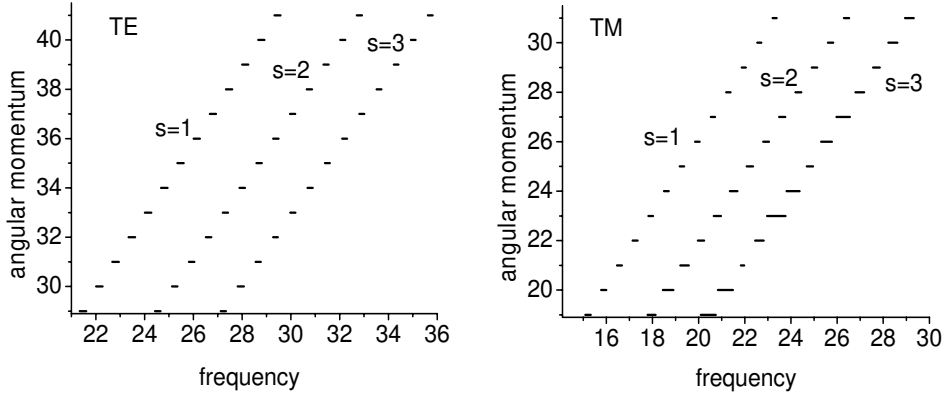}\\
  \caption{Positions and widths of several TE and TM supermodes obtained in a single band approximation. }\label{fig:band_overlap}
\end{figure}
and the equation for TM bands has the same form  with an obvious
substitution of $\beta_l$ instead of $\alpha_l$. In order to isolate
a single band, and convert this equation into a standard
tight-binding form, we can take advantage of the fact that the
expected width of the band of the $l,s$ supermode, $\delta x_{ls}$,
is rather small, $\delta x_{ls}\ll x_{ls}$. In this case we can
expand the single sphere scattering parameters $\alpha$ or $\beta$
around their respective resonance frequencies $x_{ls,TE}$ or
$x_{ls,TM}$. Taking into account the resonant nature of the
scattering parameters, we can write down:
\begin{equation}\label{eq:alpha_expen}
    \dfrac{1}{\alpha_{l}}\approx\dfrac{\left(  x-x_{ls}+i\gamma_{ls}\right)
}{i\Delta_{ls}},
\end{equation}
and the similar equation for the TM polarization. Parameters
$\Delta_{ls}$ in Eq.(\ref{eq:alpha_expen}) are real valued positive
quantities for all $l$ and $s$\cite{MiyazakiPRB2000}, and
$\gamma_{ls}$ represents the rate of the radiative decay of the
respective WGM. Taking into account that translation parameters
$A_l^{l^\prime}$ in the maximum term approximation are almost purely
imaginary quantities:
\begin{equation}\label{eq:A_abs}
    A_{l}^{l^\prime}\cong i\left(  -1\right)  ^{l+1}\left\vert A_{l}^{l^\prime}\right\vert
\end{equation}
and calculating them at the resonance frequencies $x_{ls}$, one can
re-write Eq.(\ref{eq:TE_SM}) or similar equation for the TM
polarization in the standard for the tight-binding approximation
form
\begin{equation}\label{eq:SMTB_standard}
\left( x-x_{ls}+i\gamma_{ls}\right)a_{l}^{i}=\left( -1\right)
^{l}\Delta_{ls}\tilde{A}_{ls}\left(a_{l}^{i-1} +a_{l}^{i+1}\right),
\end{equation}
where $\tilde{A}_{ls}=\left\vert
A_{l}^{l}\left(x_{ls}\right)\right\vert$. Normal modes of this
system of equations are harmonic waves
\begin{equation}\label{eq:SMnormal modes}
    a_l^i\propto \exp(iq_{ls}z_i)
\end{equation}
where $z_i$ is the $z$ coordinate of the $i$-th sphere, and
$q_{ls}(x)$ is a Bloch wave number. It satisfies the dispersion
equation of the same type as Eq.(\ref{eq:SMTBMdisp_eq}), where we
can now give a microscopic expression for the phenomenological
coupling parameter $\kappa$:
\begin{equation}
\kappa_{ls}=(-1)^{l}2\Delta_{ls}\tilde{A}_{ls}
\end{equation}

Dispersion equation (Eq.(\ref{eq:SMTBMdisp_eq})) describes a band of
excitations, which is symmetric with respect to the WPG frequency
$x_{ls}$, and whose boundaries $x_b^\pm$ are given by expression
\begin{equation}\label{eq:boundaries}
x_b^\pm=x_{ls}\pm\kappa_{ls}
\end{equation}
In the case of a chain consisting of a finite number of spheres,
$N$, the dispersion equation~(\ref{eq:SMTBMdisp_eq}) can be used to
find the spectrum of the respective frequencies. To this end, one
has to impose obvious boundary conditions for the coefficients
$a_l^i$, which read as $a_l^{0}=a_l^{N+1}=0$, where we assume that
first sphere in the chain is assigned number $i=1$. These boundary
conditions determine the allowed values of the Bloch wave number
$q_{ls}$: $q_{ls}\eta=\pi n/(N+1)$, where $n$ changes from $1$ to
$N$. Applying this result to the case of only two spheres we find
that there are two possible values of $q_{ls}$, namely
$q_{ls}\eta=\pi/3$ and $q_{ls}\eta=2\pi/3$. The respective values of
frequencies $x^{1,2}_{ls}=x_{ls}\pm \kappa_{ls}/2$, reproduce the
results of Ref.\onlinecite{MiyazakiPRB2000} obtained for the case of
a bi-sphere in the single mode approximation. The radiative decay
characterized by $\gamma_{ls}$ makes the supermodes
quasi-stationary; obviously the whole concept of the collective
excitations with different $q$ can only make sense if
$\gamma_{ls}\ll\kappa_{ls}$. Calculations as well as
experiments\cite{MoellerOL2005,MukaiyamaPRL2005,AstratovAPL2004}
show that for modes with large $l$ this inequality is well
satisfied. 

This simple picture could describe supermodes that do not overlap
spectrally since only in this case omitting terms with
$l^\prime\ne l$ in Eq.(\ref{eq:TE_NNA}) can be justified. The
bands of collective excitations in the chain of microspheres,
however, almost certainly overlap with at least one or more other
bands. In Fig.\ref{fig:band_overlap} we present positions and
widths of a number of supermode bands calculated in the single
band approximation. It should be understood, however, that because
of the inherently present radiative decay, the concept of allowed
and forbidden bands is not very well defined even for modes with
relatively large Q-factors. Nevertheless, we will use these terms
to describe spectral regions, where the wave number $q$ would have
been purely real or purely imaginary in the absence of the decay,
respectively. In reality, the wave number is a complex valued
quantity at all real frequencies, and a difference between allowed
and forbidden bands manifests itself only in relative magnitudes
of real and imaginary parts of $q$:
$\operatorname*{Im}q\ll\operatorname*{Re}q$ within the allowed
band, and $\operatorname*{Im}q\gg\operatorname*{Re}q(\mod\pi)$
within the forbidden bands. In the vicinity of the band
boundaries, $x_b^\pm$, the real and imaginary parts are of the
same order:
$\operatorname*{Im}q\simeq\operatorname*{Re}q$. 

\subsection{Inter-band coupling and band hybridization in the linear
chain of spheres}\label{sec:coupling} In this subsection we will
take into account non-diagonal  terms in Eq.(\ref{eq:TE_NNA}) and
(\ref{eq:TM_NNA}), which are responsible for coupling between bands
with different angular momentum described in the previous
subsection. Since calculations for TE and TM modes are almost
identical, we shall consider explicitly only TE polarization;
results for TM polarization, which we will use for numerical
examples, can be easily obtained with the help of obvious
substitution of the parameters.

A particular solution to the system of equations (\ref{eq:TE_NNA})
can be easily found in the form of a harmonic wave:
$a_l^{(i)}=a_le^{iqz_i}$, where $a_l$ satisfy the following
algebraic equations:
\begin{equation}\label{eq:TEamplitudes_general}
    \frac{1}{\alpha_l}a_l+2\cos\left( q\eta\right)\sum_{l^\prime}A_l^{l^\prime}a_{l^\prime}=0
\end{equation}
Let us consider this equation in a frequency region occupied by a
single-band supermode $l_0TEs_0$ characterized by quantum numbers
$l=l_0$ and $s=s_0$. We are interested in finding corrections to
the single band dispersion law for this supermode caused by its
interaction with other bands. In order to develop an approximate
analytical solution for this problem we have to take into account
several circumstances. First, main corrections to the diagonal
approximation come from those $l$, which correspond to bands
spectrally overlapping with $l_0TEs_0$. It should be remembered,
however, that the bands discussed in the previous subsection are
broadened because of the radiative decay, and the term overlap
should be understood with this fact in mind. One can see from
Fig.\ref{fig:band_overlap} that all supermodes with $l>l_0$ are
spectrally well separated from $l_0TEs_0$, thus the respective
terms in Eq.(\ref{eq:TE_NNA}) can be considered a small
perturbation despite the fact that $A_{l_0}^l>A_l^l$ for $l>l_0$.
Terms with $l<l_0$ are more important since according to
Fig.\ref{fig:band_overlap} respective bands can be spectrally in
close proximity to $l_0TEs_0$, or even overlap with it. On the
other hand coupling coefficients $A_{l_0}^{l}$ quickly decrease
when $l$ becomes smaller than $l_0$. Thus, the effect of coupling
to these supermodes depends on an interplay between resonant
enhancement due to spectral proximity, and decrease in the
coupling parameter $A_{l_0}^{l}$.

In most practical situation, among all the bands contributing to
the sum over $l^\prime$  there is just one, which we will label as
$l_1TEs_1$, effecting the supermode under consideration in a most
significant way. Usually the interaction with such a band is too
strong to allow for a perturbative treatment. Following the
terminology of Ref.\onlinecite{MiyazakiPRB2000} we will call
$l_0TEs_0$  and $l_1TEs_1$ significant supermodes, while the
remaining ones are called bath or reservoir bands and can be
treated perturbationally. We will show here that in the system
under consideration the inter-band coupling can result in three
qualitatively different types of the modifications of the
dispersion laws. One, which is characteristic for weaker
interaction can be called band renormalization, while the others,
requiring a stronger coupling, are more appropriately called band
hybridization. There might be two types of the hybridization, one
arising, when the interacting bands are characterized by angular
momentum indexes of the same parity, and the other one
corresponding to the situation,  when they have angular momentum
indexes of different parities. We will show that the hybrid band
arising in each of these cases exhibit qualitatively different
dispersion properties.

We will begin by developing a general theory capable of describing
all possible situations, and then consider conditions controlling
a transition from one regime to the other. As it was mentioned
before, the interaction between $l_0TEs_0$ and $l_1TEs_1$ should
be taken into account exactly, while the contribution from all
other
supermodes can be treated perturbatively. 
Accordingly, we  present the system of
Eq.(\ref{eq:TEamplitudes_general}) for the amplitudes of the
supermodes in the following form:
\begin{equation}\label{eq:two_mode}
    \begin{split}
&\left[\frac{1}{\alpha_{l_0}}+2A_{l_0}^{l_0}\cos\left(
q\eta\right)\right]a_{l_0}+2\cos\left(
q\eta\right)A_{l_0}^{l_1}a_{l_1}+2\cos \left(q\eta\right)\sum_{l\ne
l_0,l_1 }A_{l_0}^{l}a_{l}=0\\
&\left[\frac{1}{\alpha_{l_1}}+2A_{l_1}^{l_1}\cos\left(
q\eta\right)\right]a_{l_1}+2\cos
\left(q\eta\right)A_{l_1}^{l^0}a_{l^0}+2\cos
\left(q\eta\right)\sum_{l\ne
l_0,l_1 }A_{l_1}^{l}a_{l}=0\\
&\left[\frac{1}{\alpha_{l}}+2A_l^l\cos\left(
q\eta\right)\right]a_{l}+2\cos \left(q\eta\right)\sum_{l^{\prime}\ne
l }A_{l}^{l^\prime}a_{l^\prime}=0.
\end{split}
\end{equation}
where in the first two lines we extracted from the sum over $l$ the
terms with $l=l_0$ and $l=l_1$, and wrote down the separate
equations for the respective amplitudes. The last line in
Eq.~(\ref{eq:two_mode}) represents the equation for the reservoir
modes, which we solve for $a_l$ and substitute the solution to the
first two lines of this equation. In the resulting double sum over
$l$ and $l^\prime$, we keep only  terms with $l^\prime=l_0$ and
$l^\prime=l_1$. This procedure, which constitutes the second order
perturbation theory for the reservoir bands, results in a system of
equations for the amplitudes of the significant supermodes, which
can be presented in the following form:
\begin{equation}\label{eq:two_mode_approx}
    \begin{split}
&\left[\frac{1}{\alpha_{l_0}}+2\tilde{A}_{l_0}^{l_0}\cos\left(
q\eta\right)\right]a_{l_0}+2\cos\left(
q\eta\right)\tilde{A}_{l_0}^{l_1}a_{l_1}=0\\
&\left[\frac{1}{\alpha_{l_1}}+2\tilde{A}_{l_1}^{l_1}\cos\left(
q\eta\right)\right]a_{l_1}+2\cos
\left(q\eta\right)\tilde{A}_{l_1}^{l^0}a_{l^0}=0
\end{split}
\end{equation}
The weak interaction with the reservoir results here in the
renormalization of the coupling coefficients according to the rule:
\begin{equation}\label{eq:A_renorm}
A_l^{l^\prime}\rightarrow
\tilde{A}_l^{l^\prime}=A_l^{l^\prime}\left[1+\cos
\left(q\eta\right)\sum_{\nu\ne
l,l^\prime}\frac{\Omega_{l\nu}^{l^\prime}}{\cos\left(
q\eta\right)-\cos\left( q_\nu\eta\right)}\right],
\end{equation}
where
\begin{equation}\label{eq:Omega}
  \Omega_{l\nu}^{l^\prime}=  \frac{A_{l}^{\nu}A_\nu^{l^\prime}}{A_l^{l\prime}A_\nu^\nu}
\end{equation}
is the inter-band interaction constant, and $\cos\left(
q_\nu\eta\right)$ corresponds to the single band approximation for
the reservoir modes defined as
\begin{equation}\label{eq:zero_approxim_disp_law} \cos\left(
q_\nu\eta\right)=-\frac{1}{2\alpha_\nu A_\nu^\nu}.
\end{equation}
In contrast to Sec.\ref{sec:single_band} we have omitted here
subindex $s$ in  our notations for wave numbers $q_{\nu}$. The
reason for this is that there is no summation over $s$ in any of the
equations describing the inter-supermode coupling, such as
Eq.~(\ref{eq:TEamplitudes_general}) or (\ref{eq:two_mode}), and the
dependence on this index appears in an explicit form only if we
expand scattering coefficients $\alpha_l$ around a respective WGM.
Unlike Sec.\ref{sec:single_band} we cannot do this here, because we
take into account contributions from bands in the frequency region,
which can be relatively far away from their respective parent WGMs.
Thus, the sum over $\nu$ in Eq.~(\ref{eq:A_renorm}) contains
contributions from terms for which the vicinities of $x_{l_0s_0}$
and $x_{l_1s_1}$ belong to the forbidden bands of respective
supermodes, so that their (mostly imaginary) parameters $q_\nu$ ,
defined by Eq.~(\ref{eq:zero_approxim_disp_law}), can no longer be
associated with a particular WGM.

Eq.(\ref{eq:two_mode_approx}) describes new  bands formed from the
initial $l_0TEs_0$ and $l_1TEs_1$ bands. In order to simplify
notations we will omit for now the angular momentum indices, and
will label these new bands simply as $q_+$ and $q_-$.   We can
present dispersion equations for each of $q_\pm$ in the form:
\begin{equation}\label{eq:disp_eq_exact}
\begin{split}
\cos( q_{\pm}\eta)& = \dfrac{1}{2} \left(
\dfrac{1}{1-\tilde{\Omega}_{l_0l_1} } \right)\times \\
&\left[  \cos( \tilde{q}_{l_0}\eta) +\cos(\tilde{q}_{l_1}\eta)
\pm\sqrt{\left[ \cos(\tilde{q}_{l_0}\eta) -\cos(\tilde{q}_{l_1}\eta)
\right] ^{2} +4\tilde{\Omega}_{l_0l_1} \cos(\tilde{q}_{l_0}\eta)
\cos(\tilde{q}_{l_1}\eta) } \right],
\end{split}
\end{equation}
where $\tilde{\Omega}_{l_0l_1}$ is defined by Eq.(\ref{eq:Omega}),
in which we set $l^\prime=l=l_0$, $\nu=l_1$, drop the upper index as
duplicate, and replace the coupling coefficients with their
renormalized according to Eq.(\ref{eq:A_renorm}) values. Similarly,
$\tilde{q}_l$ are defined as in
Eq.(\ref{eq:zero_approxim_disp_law}), but also with renormalized
coupling coefficients.

Eq.(\ref{eq:disp_eq_exact}) can be considered either as equations
for wave numbers $q_\pm$ as functions of frequency $x$, or as
equations for two frequencies $x_\pm$ as functions of wave number
$q$. The choice depends on an experimental situation under
consideration. In the transport experiments of the type carried out
in Ref.~\onlinecite{MoellerOL2005} or \onlinecite{AstratovAPL2004}
the frequency is fixed by an external excitation, and $q$ is
determined from the experiment. In this case we have to solve this
equation for $q_\pm$ treating $x$ as an independent real valued
variable. The resulting wave numbers are complex, and their
imaginary parts characterize the spatial decay of the wave along the
chain. On the other hand, in the resonance experiments of
Ref.~\onlinecite{MukaiyamaPRL2005} the wave number is fixed, and
frequency is measured. In this case, we have to solve our dispersion
equation for $x$ considering $q$ as a real quantity. The frequencies
obtained as a result contain imaginary parts, which describe the
spectral width of the respective resonances. It is important to
understand that because the dispersion equation is complex, the
functions $Re(q_\pm(x))$ and $Re(x_\pm(q))$ obtained in these two
approaches are not inverse of each other. In this paper we will
focus on finding $q_\pm(x)$ because it presents greater interest
from the stand point of experiment as well as applications.

For two special values of frequency, Eq.(\ref{eq:disp_eq_exact}) can
be solved exactly. Indeed, consider $x=x_{l_0s_0}$, where
$\cos(\tilde{q}_{l_0})$ is exactly equal to zero regardless of the
renormalization of the coupling coefficients. In this case the term
responsible for the inter-band coupling vanishes, and we obtain that
one of the $\cos q_\pm$ is also equal to zero. The same is obviously
valid for $x=x_{l_1s_1}$. We can conclude, thus that the centers of
all bands, where $q_\pm\eta=\pi/2$, correspond to frequencies of the
respective parent WGM resonances regardless of the presence of the
inter-band coupling. In resonant experiments with finite
chains\cite{MukaiyamaPRL2005} the admissible values of $q$ are
determined by the boundary conditions at the ends of the chain. In
this situation, the center of the band is accessible only in systems
with odd number of spheres. Therefore, spectra of chains with odd
and even number of spheres can be distinguished from each other by,
respectively, presence or absence of  excitations at frequencies
corresponding to single sphere whispering gallery modes. This
conclusion is in complete agreement with observations of
Ref.\onlinecite{MukaiyamaPRL2005}.

In general case we can solve Eq.(\ref{eq:disp_eq_exact}) numerically
by consecutive iterations. The zero iteration corresponds to
neglecting the bath-induced renormalization of the coupling
coefficients, and produces two zero-order dispersion curves
$q_\pm^{(0)}$. At the next step, $q_\pm^{(0)}$ are substituted to
Eq.~(\ref{eq:A_renorm}) and Eq.~(\ref{eq:zero_approxim_disp_law})
producing a pair (one for $q_+^{(0)}$ and one for $q_-^{(0)}$) of
new values of the coupling coefficients and "single-band" wave
numbers $\tilde{q}_l$. These new values go back to
Eq.(\ref{eq:disp_eq_exact}), one to the "$+$" version of it, and the
other to the "$-$" version. The procedure can be repeated as many
times as necessary to achieve its convergence. The experience shows,
however, that  a good approximation for $q_\pm$ can already be
obtained  after only the first iteration.

While the contributions from the reservoir bands can be important,
the main qualitative characteristics of the dispersion laws of the
significant supermodes can be understood from the zero order
approximation, which is presented by Eq.(\ref{eq:disp_eq_exact})
without the renormalization of the coupling coefficients. We are
focused on the frequency region, where expression $\cos q_{l_0}
-\cos q_{l_1}$ is small because this is where the main effects of
the inter-band interaction are expected. The strongest effects occur
in the vicinity of a resonance between supermodes $l_0$ and $l_1$,
when
\begin{equation}\label{eq:reson}
\operatorname*{Re}\left[\cos\left(
q_{l_0}\eta\right)\right]=\operatorname*{Re}\left[\cos\left(
q_{l_1}\eta\right)\right],
\end{equation}
if it takes place in the frequency range covering initial
$l_0TEs_0$ and $l_1TEs_1$ bands. It should be noted, however, that
because of the radiative decay the band boundaries particularly
for the band with larger $s$ are not that very well defined, and
the strong effects can take place even if the resonance condition
is fulfilled in the forbidden band of one or even both of the
modes, if the resonance point lies in proximity of a band
boundary. It is important to remember also that
Eq.(\ref{eq:reson}) is not equivalent to $\operatorname*{Re}(
q_{l_0})=\operatorname*{Re}( q_{l_1})$, which is often accepted as
a resonance condition.

The properties of the dispersion curves in the vicinity of the
resonance point are determined by two circumstances: (i) the
parity of single band dispersion laws, and (ii) the relation
between $[\operatorname*{Im}\left[\cos\left(
q_{l_1}\eta\right)\right]-\operatorname*{Im}\left[\cos\left(
q_{l_0}\eta\right)\right]]^2$ and $\Omega_{l_0l_1}
\cos(q_{l_0}\eta) \cos(q_{l_1}\eta)$, which determines if the band
interaction is weak or strong. In most cases the overlapping
supermodes originate from WGM with different $s$, and, therefore,
usually, one of the imaginary parts in the expression above is
significantly (orders of magnitude) larger than the other one. In
this case the stronger decaying supermode determines the nature of
the modification of the dispersion laws. It is quite obvious that
the strong coupling case (band hybridization) would correspond to
the interaction terms prevailing over the dissipative terms, and
the weak coupling regime (band modification) would take place in
the opposite situation. If the resonance condition is not
fulfilled in the spectral region of interest the modification of
the dispersion law always belongs to the weak coupling case. What
is surprising, however, is that even in this case the modification
can be so strong as to render the perturbative treatment of
interaction between the significant modes not very accurate.

An  important peculiarity of the dispersion curves described by
Eq.(\ref{eq:disp_eq_exact}) distinguishing it from other cases of
mode coupling (polaritons, for instance) is the sign of the
inter-band interaction parameter, $\Omega_{l_0l_1}$. Using
properties of the coupling coefficients $A_l^{l^\prime}$ it is easy
to show that this parameter is always positive regardless of the
parities of the angular momentum indexes $l$ and $l^{\prime}$. The
parity of $l$ according to Eq.(\ref{eq:SMTB_standard}) determines
the sign of the slope of the respective dispersion laws in the
vicinity of band boundaries. Taking into account that in the
frequency region of interest $\cos(q_{l_0}\eta)$ and $
\cos(q_{l_1}\eta)$ are, at least, of the same sign, we conclude that
the entire interaction term $\Omega_{l_0l_1} \cos(q_{l_0}\eta)
\cos(q_{l_1}\eta)$ is always positive regardless of the interacting
bands having slopes of the same or different signs.  We will see in
the subsequent sections that this feature of the inter-band
interaction, which results from the non-Hermitian nature of the
matrix $A_l^{l^\prime}$, is responsible for a rather unusual form of
the dispersion laws describing hybridized bands.

\subsubsection{Weak inter-band coupling}\label{subsec:weak_coupling}
\begin{figure}
  \includegraphics[width=5in]{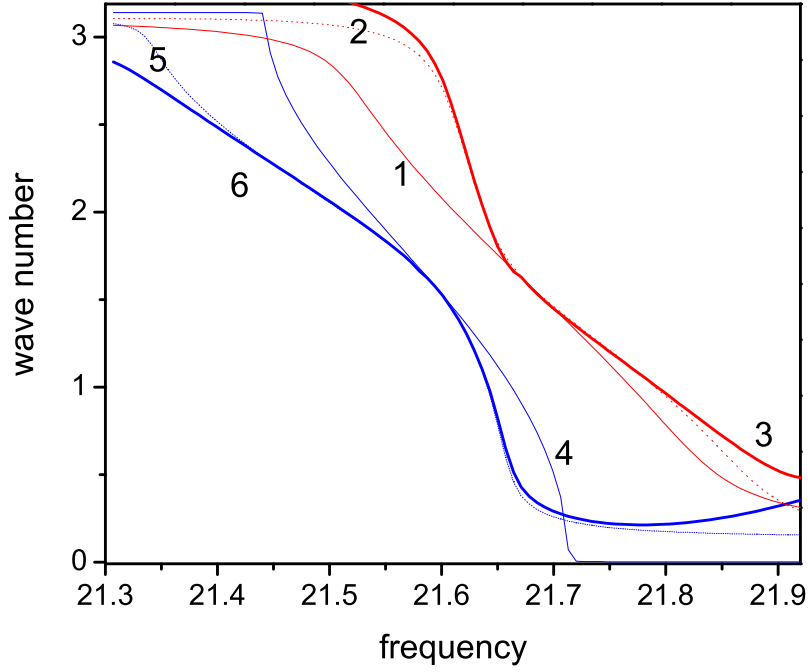}\\
\caption{Dispersion curves of $29TM1$ and $24TM2$ bands. Dash lines
labelled $1$ and $4$ presents single-band dispersion curves, thin
lines labelled $2$ and $5$ demonstrates dispersion cruves found
without reservoir modes taken into account, and finally thick lines
labelled $3$ and $6$ show dispersion curves with reservoir modes
included  }\label{fig:TM291&TM252disp_eq}
\end{figure}
As an example of the band modification we consider dispersion curves
emerging as a result of interaction between $29TM1$ and $25TM2$
supermodes. This choice is motivated by experiments of
Ref.\onlinecite{MukaiyamaPRL2005}, where these dispersion curves
have been measured. One can see from Fig.\ref{fig:band_overlap} that
the bands of these two supermodes slightly overlap. However,
considering $\operatorname*{Re}\left[\cos\left(
q_{29}\eta\right)\right]$ and $\operatorname*{Re}\left[\cos\left(
q_{25}\eta\right)\right]$ we find that the resonance between these
two bands does not occur, thus, we should expect in this case just a
modification of the dispersion laws without a significant
reconstruction of the band structure.

Fig.\ref{fig:TM291&TM252disp_eq} presents the results obtained after
first two iterations of numerical solutions of
Eq.(\ref{eq:disp_eq_exact}) for these two modes along with the
respective initial bands. The second iteration takes into account
all reservoir modes from $l=1$ to $l=40$. One can see that the
effect of the reservoir modes is almost zero in the vicinity of the
center of the band, but slightly increases toward the edges of the
initial bands, which confirms our choice of significant and
reservoir modes. However, outside of the allowed bands of the
$29TM1$ and $TM24,2$ modes designation of these modes as significant
ones may no longer be valid because we can trespass to the allowed
band of one or several of other modes, which we treated here as the
bath. It is obvious that in those frequency regions the
approximation used to obtain Fig.\ref{fig:TM291&TM252disp_eq} is no
longer valid.

\begin{figure}
  \includegraphics[width=5in]{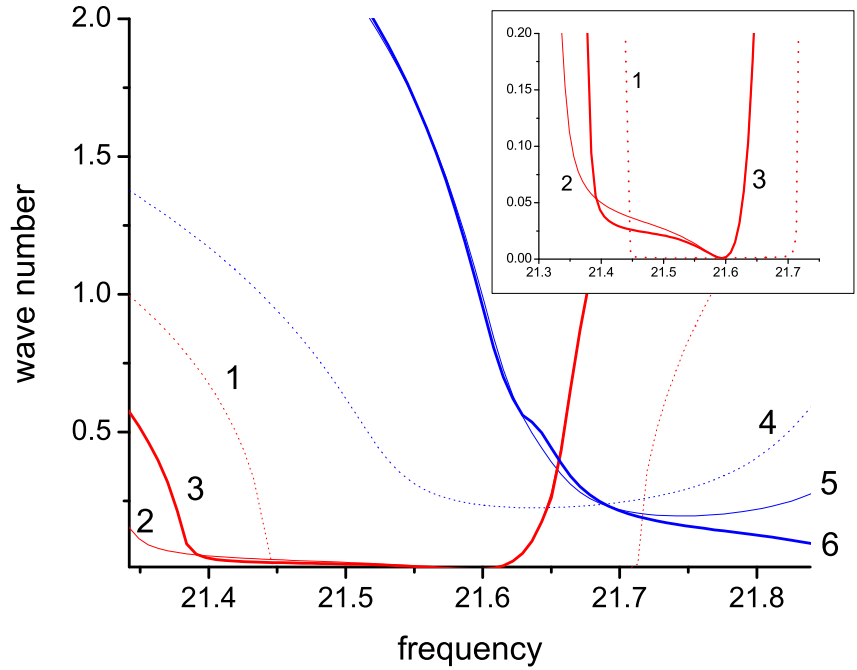}\\
\caption{Decay rates of $29TM1$ ($1$, $2$ and $3$) and $24TM2$ ($4$,
$5$, and $6$) bands. The labelling of the curves is the same as in
Fig.\ref{fig:TM291&TM252disp_eq}. The insert shows, in a magnified
form, the spectral region corresponding to 29TM1 mode. One can see a
significant increase in the radiative decay rate of this mode in the
vicinity of the new band boundaries, which, however, decreases to
the initial WGM value at $x=x_{29,1}$}\label{fig:TM291&TM252decay}
\end{figure}

The shape of the modified curves can be easily understood on the
basis of our general analysis of Eq.(\ref{eq:disp_eq_exact}). As we
explained above at $\tilde{q_{l}}\eta=\pi/2$ modified and initial
dispersion laws coincide, but farther  away from the center of the
respective bands the modified dispersion laws are pushed away from
the initial curves. The characteristic shape of the dispersion
curves results from the fact that the corrections to the initial
dispersion laws due to the inter-band coupling have different signs
for  $29TM1$ and $25TM2$ bands: its negative for the former and
positive for the latter.  The most strong modifications of both
curves occur at the lower frequency boundaries of their respective
bands. This modification can be described as a shift of the
boundaries of the respective bands, which is more prominent for
$29TM1$, whose initial band has much better defined boundaries
because of the smaller decay. These boundaries are not very well
defined for modified $29TM1$ band, however,  which tells us about an
increase in the radiative decay of this supermode caused by the
admixture of $25TM2$ band. This increase, however, is not very
dramatic, and allows keeping the concept of allowed and forbidden
bands as a convenient tool to characterize different spectral
regions. The frequency dependence of the decay rates, characterized
by $\operatorname*{Im}\tilde{q}_l$ are shown in
Fig.\ref{fig:TM291&TM252decay}, where the positions of the band
boundaries characterized by an abrupt change in the decay rate are
seen much clearer.
Another interesting effect seen in this figure is that the
frequency $x_{29TM1}$ of the respective WGM is no longer at the
center of the modified band. This is, obviously, caused  by the
shift of the band boundaries, which for both low- and
high-frequency edges occurs toward lower frequencies. This shift
is also responsible for an asymmetry of the band, which manifests
itself in the different growth of the decay rates toward left and
right boundaries.

We can estimate the shifts of the band boundaries for $29TM1$ and
$25TM2$ supermodes if we neglect the renormalization of the
coefficients $A_l^{l_1}$, and treat the inter-band interaction
parameter $\Omega_{l_1l_0}$ as a small perturbation. Then,
employing the approximation for coefficients $\alpha_l$ given by
Eq.(\ref{eq:alpha_expen}), and neglecting the frequency dependence
of the coupling coefficients we can find for new band boundaries,
$\tilde{x}_{b}^\pm$, defined as frequencies at which
$\operatorname*{Re}[\cos(q_\pm\eta)]=1$ the following expressions
\begin{equation}\label{eq:new_bound}
\tilde{x}_{b_l}^\pm=x_{b_l}^\pm\mp\frac{\Omega_{l\nu}\kappa_{ls}}{1\mp\cos\left[q_\nu\left(x_{b_l}^\pm\right)\eta\right]}
\end{equation}
Index $l$ in this equation refers to the band whose boundaries we
are calculating, while the index $\nu$ signifies its interacting
counterpart. Since $\Omega_{l\nu}$ is always positive, and
$\kappa_{ls}$ is negative for both participating modes, the sign
of the correction  is determined by the values of $\cos q$ of the
$\nu$-th band at the initial boundaries of the $l$-th band. Since
its quite likely that these boundaries lie in the forbidden band
of the $\nu$-th band, the values of these cosines are not limited
by unity. In the case of $29TM1$ supermode interacting with
$25TM2$ we find that
$$
\cos\left[q_{25}(x^+_{29,1})\eta\right]\approx 0.18 \hspace {2 cm}
\cos\left[q_{25}(x^-_{29,1})\eta\right]\approx -1.46
$$
$$
\cos\left[q_{29}(x^+_{25,2})\eta\right]\approx 2.53\hspace{2 cm}
\cos\left[q_{29}(x^-_{25,2})\eta\right]\approx -0.5174
$$
These calculations show that the corrections to the upper and
lower boundaries are both negative for $29TM1$ band, and both
positive for $25TM2$ band in complete agreement with numerical
calculations.

\begin{figure}
  \includegraphics[width=5in]{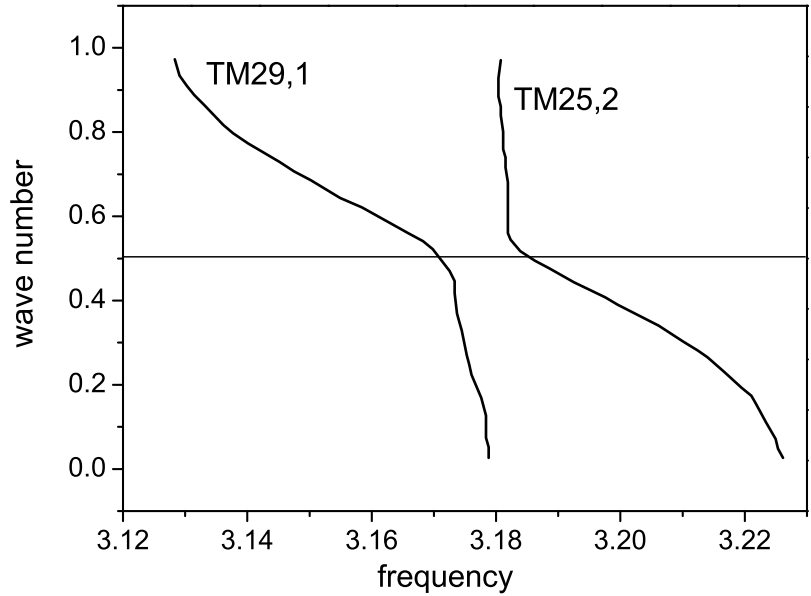}\\
  \caption{Experimental dispersion curves of $29TM1$ and $TM24,2$ bands obtained by digitizing one of the figures in Ref.\protect\onlinecite{MukaiyamaPRL2005}. }\label{fig:experiment}
\end{figure}

As it was mentioned we choose the supermodes $29TM1$ and $25TM2$ for
illustration of our general results mostly because their dispersion
curves  were observed experimentally in
Ref.\onlinecite{MukaiyamaPRL2005}. However, before comparing the
experimental results with our theory we should recall that the
experiment in Ref.\onlinecite{MukaiyamaPRL2005} was conducted in a
resonance configuration, when $q_l$ was fixed by the conditions of
the experiment, and real and imaginary parts of frequency were
measured via positions and widths of the respective resonances. This
experimental setup is different from what was assumed in our
theoretical analysis. Therefore, the comparison makes sense only
away from the band boundaries, where decay rates are relatively
small and the inverse of the experimentally observed quantity
$x_l(q)$ is close to $q_l(x)$ studied here. Fig.\ref{fig:experiment}
demonstrates the curves $q_{29,1}(x)$ and $q_{25,2}(x)$ produced by
digitizing Fig.2b from Ref.\onlinecite{MukaiyamaPRL2005}. Comparing
this figure with results of our calculations,
Fig.\ref{fig:TM291&TM252disp_eq} one can see an excellent
qualitative agreement between the theory and the experiment.
\subsubsection{Strong inter-band coupling: band hybridization}
\begin{figure}
  \includegraphics[width=5in]{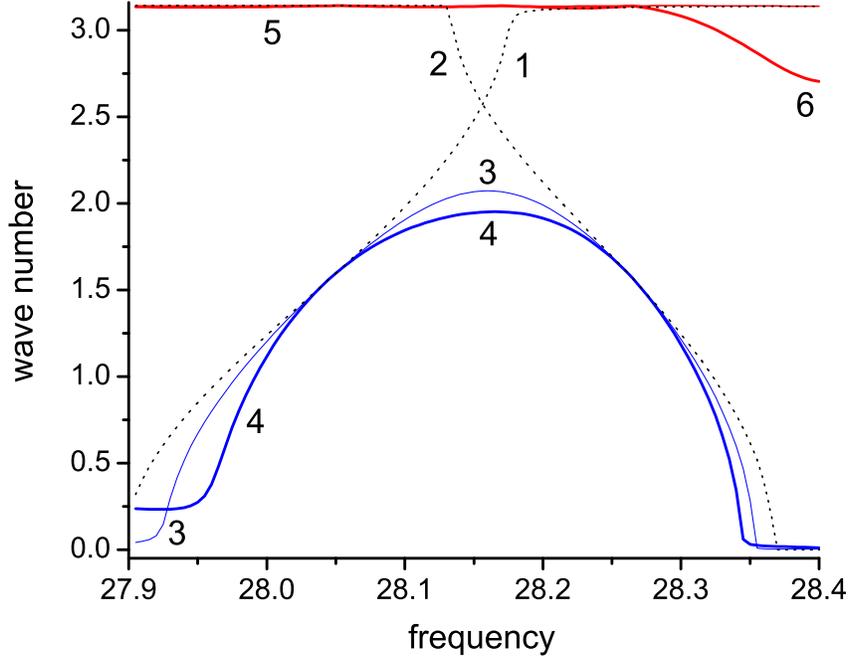}\\
  \caption{Dispersion laws describing hybridized bands arising as a result of interaction between $39TE1$ and $34TE2$ supermodes.
  Curves 1 and 2 represents initial unperturbed dispersion laws, 3 and 5 are dispersion curves obtained without accounting for interaction with reservoir modes,
  and curves 4 and 6 depict dispersion laws of two branches with reservoir modes taken into account. The upper branch is almost completely pushed to the forbidden region.      }\label{fig:TE391&TE342disp_eq}
\end{figure}

In order to illustrate the effects of the strong inter-band coupling
we consider the interaction between modes $39TE1$ and $34TE2$. The
resonance condition (\ref{eq:reson}) for these modes is fulfilled at
frequency $x_r$, which is within allowed spectral region for both
bands, albeit rather close to their respective band boundaries.
Therefore, we should expect the manifestation of the strong coupling
effects in this case. Another important feature distinguishing this
pair of modes is the different parity of their angular momentum
indexes. As a result, the initial (single-band) dispersion curves of
these supermodes are characterized in the vicinity of the resonance
point by slopes of opposite signs. In the case of regular
interacting modes such as phonon- or exciton-polaritons, one would
expect in a situation like this a normal anti-crossing behavior
resulting in the opening of the polariton band-gap in the spectrum
of the system. This would have happened in the case considered here
as well, if the interaction term in Eq.(\ref{eq:disp_eq_exact}) were
negative. It, however, is always positive as it was explained above
and, therefore, no spectral gap arises. The shape of the dispersion
curves emerging in this situation and shown in
Fig.\ref{fig:TE391&TE342disp_eq} is rather unusual. The parts of the
initial dispersion curves located below the crossing point are
pushed downward from it and connect in a continuous manner to form a
new hybrid band. The remaining parts of the initial dispersion laws
are pushed upward such that the wave numbers corresponding to the
resultant second dispersion curve become complex valued in the
entire spectral region under consideration, with its real part
remaining equal to $\pi$ for all considered frequencies. In other
words, the spectral region covering initial $39TE1$ and $34TE2$
bands is a forbidden band for the second dispersion curve. Thus, the
inter-band coupling turns two co-existing initial bands into a
single hybrid band with rather unusual dispersion properties.
Instead of a spectral gap we have here a region of forbidden values
of \emph{wave numbers}, which extends from the maximum value of the
wave number on the lower dispersion curve, $q_{max}$, all the way to
the boundary of the Brillouin zone. This means that the states
characterized by the wave numbers from the forbidden region cannot
have a real valued frequency, i.e. do not correspond to stationary
solutions of the respective dynamic equations.

The new allowed hybrid band is a combination of the initial bands
for the two interacting supermodes and covers a spectral region
previously occupied by $39TE1$ and $34TE2$. The shifts of the
initial band boundaries (the low frequency boundary for $34TE2$ and
higher frequency boundary for $39TE1$) can be estimated with the
help of the same expression (\ref{eq:new_bound}) that we obtained
for the weak coupling case. This is possible because these band
boundaries lie far away from the resonance point and we can use
perturbation theory. Taking into account that the sign of parameters
$\kappa$ is now different for the two supermodes we find that the
the low frequency boundary of $34TE2$ shifts toward higher
frequencies by approximately $0.025$, while the higher frequency
boundary of $39TE1$ supermode shifts toward lower frequencies by
approximately $0.01$. These numbers are in good agreement with
numerical calculations shown in Fig.\ref{fig:TE391&TE342disp_eq}.
This figure also shows that taking into account the reservoir modes
(second iteration of our procedure) does not change properties of
the hybrid mode too much.

While the existence of the wave number gap is the general property
of the dispersion equation (\ref{eq:disp_eq_exact}) for the
significant supermodes characterized by angular momentum index of
different parity, the complete vanishing of the upper hybrid band is
specific for the pairs of modes under consideration here. In order
to understand the properties of these bands better we can estimate
the distance between the lower and upper branches exactly at the
resonance point neglecting the effects due to the reservoir modes.
This distance, $\Sigma$, determines the position of the maximum
$q_{max}$ for the lower branch, and the minimum $q_{min}$ for the
upper branch. It is more convenient to work with $\cos q_\pm$ than
with the wave numbers themselves, so we will define $\Sigma$ via
equation:
\begin{equation}\label{eq:wave_num_gap}
\begin{split}
    \cos\left(q_{max}\eta\right)=\cos\left(q_r\eta\right)+\Sigma\\
    \cos\left(q_{min}\eta\right)=\cos\left(q_r\eta\right)-\Sigma
    \end{split}
\end{equation}
The magnitude of $\Sigma$ can be estimated from
Eq.(\ref{eq:disp_eq_exact}), which gives
\begin{equation}\label{eq:Sigma}
    \Sigma=\frac{\operatorname*{Re}\left[\sqrt{4\Omega_{l_1l_0}-\Gamma^2-4i\Omega_{l_1l_0}\Gamma}\right]}{1-\Omega_{l_1l_0}}
\end{equation}
where we took into account that the resonance occurs close to the
band boundaries, so that $\operatorname*{Re}[\cos
q_{l_0}]\approx\operatorname*{Re}[\cos q_{l_1}]\approx -1$, and
introduced $\Gamma=\operatorname*{Im}[\cos q_{l_1}]$, which
represents the supermode with the largest radiative decay ($34TE2$
in the case under consideration). However, the radiative decay of
the $34TE2$ supermode is still so small that
$\Gamma^2\ll\Omega_{l_0l_1}$ in this case, and $\Sigma$ can be
approximated as
\begin{equation}\label{eq:Sigma_approx}
    \Sigma=\frac{2\sqrt{\Omega_{l_1l_0}}}{1-\Omega_{l_1l_0}}
\end{equation}
In the particular example of  $39TE1$ and $34TE2$ bands the
resonance occurs at a point, where $\cos q_r$ is so close to the
boundary value of $-1$, that $\cos\left(q_{min}\eta\right)$ becomes
less than $-1$ making the respective wave numbers for the entire
frequency region imaginary. Should the crossing point of the two
interacting supermodes lie farther away from the band boundaries,
the part of the upper hybrid band could also survive, but it would
be restricted by the frequency region in the vicinity of the
resonance frequency. The $\cos\left(q_{max}\eta\right)$ on the other
hand is pushed farther away from the boundary, so that the entire
lower band remains allowed, and the respective maximum allowed value
of $q$ can be estimated as $q_{max}\eta\approx
q_r\eta-\sqrt{2\Sigma}$. Comparison of these estimates with
numerical results shows that they give a relatively good
approximation for the respective quantities in the case under
consideration.

One of the consequences of having a gap for wave numbers, is that
formally speaking, the group velocity of the excitation described by
the hybrid dispersion law diverges at $q=q_{max}$. This form of a
dispersion curve, therefore, raises a question if it is consistent
with casuality. This question was discussed in
Ref.\onlinecite{JohnPRE2004}, where a similar form of the dispersion
curve was found in an one-dimensional resonant photonic crystal. It
was shown in that work that the wave-number gap and accompanying it
infinite group velocity do not contradict causality, if attenuation
of the respective excitation is properly accounted for. Similar
anomalous behavior of the group velocity (without the wave number
gap, of course) has been  known for a long time, for instance, in
the case of exciton-polaritons.\cite{BirmanPRL1981}  A more general
statement that abnormal behavior of group velocity should be
expected in the regions of resonant absorption in all dispersive
dielectrics was proven in Ref.\onlinecite{BoldaPRA1993}.

\begin{figure}
  \includegraphics[width=5in]{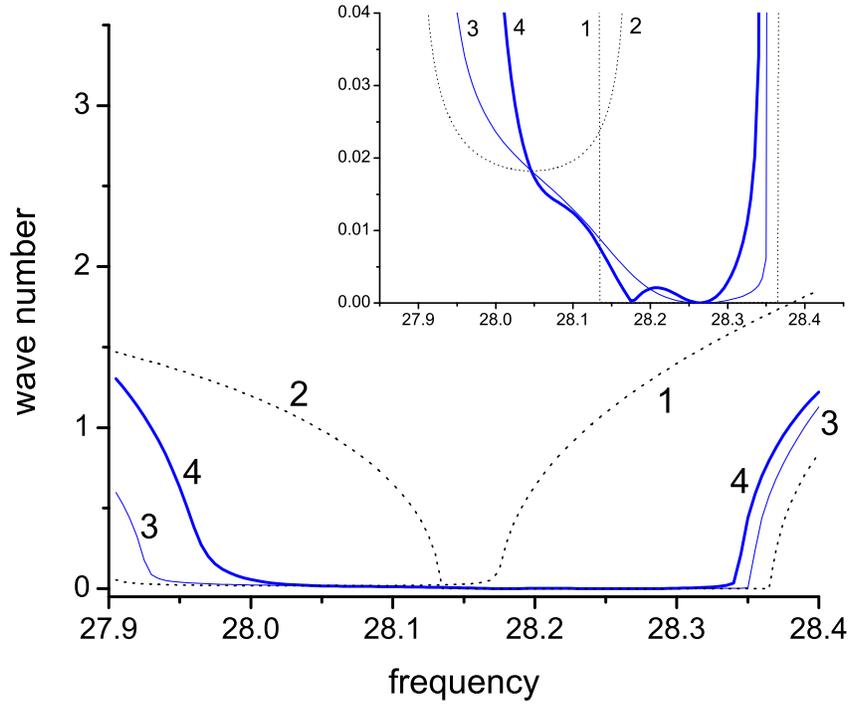}\\
  \caption{Radiative decay rates for the allowed hybridized band. The labeling of the lines is the same as in
  Fig.\ref{fig:TE391&TE342disp_eq}.
  The insert magnifies the region around the center of $39TE1$ band.}\label{fig:TE3934decay}
\end{figure}

In spite of formal analogy between the anomalous dispersion law
presented in Fig.\ref{fig:TE391&TE342disp_eq} and the results of
Ref.\onlinecite{JohnPRE2004} these two situations significantly
differ from each other. The wave number gap found in
Ref.\onlinecite{JohnPRE2004} is caused by the frequency dispersion
of the respective dielectric medium, and coincides with regions of
anomalous dispersion and resonant absorption. The origin of the gap
in our case is completely different and can be traced to the
combination of two factors: different signs of the slopes of the
initial dispersion curves and positive value of the respective
coupling constant. The latter is caused, as we already discussed, by
a non-Hermitian nature of the matrix describing the inter-band
coupling in the case under consideration. Accordingly, the decay
rate of the photonic supermodes considered here does not show any
resonant enhancement in the vicinity of the wave-number gap. We
present the frequency dependence of this quantity, which remains
surprisingly small in all frequency range corresponding to the
allowed hybrid band, and demonstrates a weak non-monotonic
dependence on frequency with a flat minimum
(Fig.\ref{fig:TE3934decay}). We can explain this behavior
qualitatively by reminding  that at the centers of both initial
bands the interaction between them vanishes, so the decay rate at
these two points coincides with the decay rates of the original WGM
resonances. The crossing of the curves 2,3 and 4 in the insert in
Fig.\ref{fig:TE3934decay} corresponds to the initial $34TE2$ WGM,
and the second (counting from the left) minimum on curve 4
corresponds to $x_{39TE1}$. The first minimum on this curve appears
only after the interaction with reservoir modes is taken into
account and presents another peculiarity of the system under
consideration. Normally, one would expect that the reservoir bands
can only increase the radiative rate, while here we observe an
opposite behavior. A possible qualitative explanation of this effect
can be offered on the basis of the following arguments. Let us
assume, for an instance, that instead of considering the
experimental situation, in which frequency is considered real and
the wave number complex, we deal with the resonance type of
experiment with real $q$ and complex $x$. In this case it is
reasonable to assume, and the results of
Ref.\onlinecite{MiyazakiPRB2000} support this assumption, that the
the reservoir modes would play its regular role and increase the
width of the respective resonances, $\Delta x$. The transition from
this description to the one used in this paper, with complex valued
$q$ can then be approximately carried out with the help of the
following expression: $\operatorname*{Im}q\simeq
(d\operatorname*{Re}q/dx)\Delta x$. Thus, the frequency regions with
flat dispersion curve should be characterized by decreasing
imaginary part of the wave number. In order to understand why this
minimum appears only after reservoir bands are taken into account,
one needs to compare respective dispersion curves. From
Fig.\ref{fig:TE391&TE342disp_eq} one can see that without the
reservoir modes the maximum of the dispersion curve is relatively
narrow. The reservoir modes make the maximum wider, which means that
a broader frequency interval is characterized by small values of
$d\operatorname*{Re}q/dx$. This causes a faster decrease of the
spatial decay rate when frequency shifts toward the allowed region
for former $39TE1$ band, which can be seen in
Fig.\ref{fig:TE3934decay}. This tendency reverses for frequencies
greater than $x_r$, when the group velocity starts decreasing. When,
however, the frequency approaches $x_{39TE1}$, the decay rate
decreases again resulting in a curve with two minima, one at $x_r$,
and the other one at $x_{39TE1}$.

\section{Conclusion}
In this paper we have developed a theory of the inter-band coupling
in the system of dielectric spheres forming a one-dimensional chain.
Our objective was to obtain analytical expressions allowing one to
study band structure and dispersion properties of the collective
excitations of this system with inter-band mixing effects taken into
account. We developed an approximation scheme generalizing an
approach of Ref.\onlinecite{MiyazakiPRB2000}, which was originally
applied to the case of two coupled spheres and derived a general
dispersion equation describing the band structure emerging as a
result of the inter-band interaction. We showed that there might
exist three qualitatively distinct regimes of such coupling
depending upon the properties of the initial single-band dispersion
curves of the interacting supermodes, and the strength of the
interaction. Our general results were applied to two particular
examples illustrating two of the possible manifestations of the
inter-mode coupling.

As an illustration of the regime of weak band modification we
considered frequency region corresponding to initial supermodes
$TE29,1$ and $25TM2$, whose initial dispersion curves do not cross
each other. Nevertheless, we showed that the inter-band coupling
significantly modifies the dispersion curves of these supermodes and
is responsible for their characteristic shape observed in
experiments of Ref.\onlinecite{MukaiyamaPRL2005}. What is
interesting in this example is that the modification of the
dispersion laws mimics the anti-crossing behavior typical for
resonantly interacting excitations. At the same time, as it was
mentioned above no crossing resonance takes place in this case. We
explained the shape of these curves as a result of two effects.
First, we showed that the position of frequency corresponding to the
center of the band, where wave number $q=\pi/(2\eta)$, where $\eta$
is dimensionless period of the structure, is pinned to the frequency
of the parent WGM mode corresponding to a given band. Second, we
demonstrated that the inter-band coupling shifts the boundary of the
respective bands in the opposite directions. As a result, one of the
dispersion curves bends upward, while the other one bends downward
imitating the anti-crossing behavior. The results of the numerical
calculations of these dispersion curves show very good qualitative
agreement with experimental results of
Ref.\onlinecite{MukaiyamaPRL2005}.

More interesting situation was found, however, in the spectral
region corresponding to supermodes $39TE1$ and $34TE2$. These modes
exhibit true crossing resonance very close to the boundaries of the
both bands. It is also important that in the vicinity of the
crossing point their initial dispersion curves have slopes of
opposite signs. The interaction between these supermodes gives rise
to a new hybrid band with a highly unusual dispersion
characteristics. Instead of a standard avoiding crossing kind of
behavior accompanied by opening of a  gap in the frequency spectrum
of the system, we found a gap in the allowed values of the
\emph{wave numbers}. More specifically, instead of two initial
crossing dispersion curves, there emerges one curve characterized by
a non-monotonic dependence of the wave number of frequency. The
maximum value of the wave number is significantly lower than
$\pi/\eta$, which determines traditional band boundaries.  Thus,
there is a range of wave numbers that do not correspond to
excitations with any real valued frequency. Such a band structure is
not related to the presence of the radiative decay, which remains
very small throughout the entire band of frequencies corresponding
to this hybrid band. We traced the origin of this effect to the
non-Hermitian nature of the inter-band coupling typical for
electromagnetic problems. The phenomenon of the wave number band-gap
could be observed experimentally in an experiment similar to one
carried out in Ref.\onlinecite{MukaiyamaPRL2005}. In that experiment
the fluorescent spectra were observed for chains with varying number
of spheres. In normal situation, adding a sphere results in
appearance of an additional peak on the spectrum so that each peak
can be identified with a respective wave number. The wave number gap
would manifest itself in this experiment as a failure of a new peak
to appear after addition of a sphere to the chain. The dependence of
the fluorescence spectrum upon the number of spheres in this
situation is a non-trivial problem requiring a separate
consideration. Other possible experimental manifestations of the
predicted band structure may include unusual behavior of a pulse
propagating along the chain, and angular dependence of the radiation
emitted by the chain. More detailed analysis of these effects  is
also outside of the scope of this paper and will be carried out in
the future.
\section*{Acknowledgments}
We would like to express our appreciation of many fruitful
discussions of this paper with A. Lisyansky, V. Menon, and E.
Narimanov. One of us (A. R.) would like to thank Vladimir Shuvaev
for help with numerical calculations.


\end{document}